\documentclass[aps,pra,showpacs,notitlepage]{revtex4-1}

\usepackage{graphicx}

\begin{document}

\title{On the relation between measurement outcomes and physical properties}

\author{Taiki Nii}
\author{Masataka Iinuma}
\author{Holger F. Hofmann}
\email{hofmann@hiroshima-u.ac.jp}
\affiliation{
Graduate School of Advanced Sciences of Matter, Hiroshima University,
Kagamiyama 1-3-1, Higashi Hiroshima 739-8530, Japan}

\begin{abstract}
One of the most difficult problems in quantum mechanics is the analysis of the measurement processes. In this paper, we point out that many of these difficulties originate from the different roles of measurement outcomes and observable quantities, which cannot simply be identified with each other. Our analysis shows that the Hilbert space formalism itself describes a fundamental separation between quantitative properties and qualitative outcomes that needs to be taken into account in an objective description of quantum measurements. We derive fundamental relations between the statistics of measurement outcomes and the values of physical quantities that explain how the objective properties of a quantum system appear in the context of different measurement interactions. Our results indicate that non-classical correlations can be understood in terms of the actual role of physical properties as quantifiable causes of the external effects observed in a quantum measurement.
\end{abstract}


\maketitle

\section{Introduction}
\label{sec:intro}

The problem with quantum mechanics is that it does not provide a satisfactory justification of the mathematical elements of the theory in terms of the actual physics happening in the laboratory. The reason for this irritating failure is that the notions used to connect the theory to the physics are hardly ever discussed. One might go so far as to say that physicists have adapted to the situation by accepting the connection between technical procedures and mathematical expressions ``on faith''. Very often, this failure to address the empirical justification of quantum theory is justified by invoking the uncertainty principle, which seems to suggest that experimental outcomes will always be random and ambiguous. However, this may not be what the uncertainty principle really means. Recently, new insights into measurement uncertainties have emerged, initiated by Ozawa's analysis of the relation between errors and disturbance in quantum measurements \cite{Oza03,Hal04,Lun10,Has12,Roz12,Bae13,Wes13,Bra13,Rin14}. In the context of these new results it is interesting to note that there is no clear consensus about the physics involved in quantum measurements \cite{Wat11,Bus13,Dre14,Bus14,Roz15}. It seems that too much of the focus has been on quantitative evaluations of the uncertainties, and too little attention has been paid to the astonishing fact that quantum mechanics does not provide a clear explanation of the measurement process itself. Recently, we have applied the procedure used in the first experimental evaluation of Ozawa's uncertainties to investigate the quantum properties of a two level system in more detail \cite{Iin16}. As we tried to understand the experimental procedure better, we found that it is necessary to re-examine the relation between the experimental evidence and the physical properties of the system. Specifically, the explanation of the Hilbert space formalism seems to be incomplete: a physical property is not defined by the outcomes of precise measurements represented by eigenstates and eigenvalues, but by quantitative effects in which different physical properties act together to produce the outcome of a specific experiment. 

In this paper, we explain how the standard formalism of quantum theory describes the relation between physical quantities and measurement outcomes. We point out that it is not possible to identify physical properties with their eigenstates and derive the correct relations between the different measurement contexts by showing that the mathematical relations between the operators provide the appropriate description of the physics of the quantum system. Most importantly, we find that the quantum formalism provides a precise description of the relations between different measurements that leaves little room for interpretation. The quantitative relations described by the operator algebra define error free assignments of quantities to specific combinations of initial condition and measurement outcomes, even when neither the initial state nor the measurement is given by an eigenstate of the physical quantity in question. In this sense, the structure of quantum theory is fully deterministic, and it is this feature of the theory that allows us to objectify measurement results obtained from different measurement procedures. 

A major concern addressed by this paper is that the one-sided focus on the statistics of quantum states may have created a widespread misconception that quantum physics is a fuzzy and random theory that is open to various interpretations. What we demonstrate in the following is a more rigorous link between experimental evidence and objective physical properties that can serve as a solid empirical foundation of quantum theory. It is our hope that this approach may bridge the gaps that have opened up between different schools of thought in quantum physics. Ultimately, real scientific questions need to be decided by the experiment, and not by the arbitrary opinions of individual researchers, no matter how qualified they may seem to be. We therefore believe that the following theory is urgently needed in order to explain the objective relations between the outcomes of different experiments, as the starting point of a more thorough and comprehensive experimental exploration of quantum physics. 

The rest of the paper is organized as follows. In Sec. \ref{sec:QM} we review the theoretical description of measurements and discuss the relation between measurement operators and the operators representing quantitative physical properties. The concept of a quantitative measurement error is introduced as a quantitative expression of this relation. In Sec. \ref{sec:jsw} the error is evaluated and its relation with individual measurement outcomes is considered. It is shown that the quantitative error can be used to assign joint statistical weights to the outcomes of two different measurements. In Sec. \ref{sec:errorfree} it is shown that error free estimates are possible whenever the target observable $\hat{A}$ can be expressed as a sum of an initial property $\hat{B}$ and a property $\hat{M}$ determined by the measurement outcomes. In Sec. \ref{sec:context} we show that the joint statistical weights introduced in Sec.  \ref{sec:jsw} explain how physical quantities depend on the specific measurement context and derive the corresponding relations between the eigenvalues of non-commuting operators. In Sec. \ref{sec:ncc} the relation between error free measurements and operator statistics is explained and non-classical correlations are identified as context independent features of quantum statistics. Finally, Sec. \ref{sec:concl} summarizes the insights gained from the quantitative analysis of measurements. 

\section{Quantum Measurements}
\label{sec:QM}

The problem of formulating a quantum theory of measurement starts with the fact that Hilbert space does not describe the system in terms of its physical properties, but provides only an indirect description in terms of state vectors, the components of which are loosely associated with potential outcomes of measurements that might or might not be performed. It is therefore insufficient to define a measurement as the observation of a pre-existing physical property. Instead, the connection between the state of a system with an eventual measurement outcome is generally expressed by a statistical operator. According to the well established rules of quantum theory, this operator $\hat{E}_m$  defines the probability $P(m|\psi)$ of obtaining the outcome $m$ as a product trace of the operator with the density operator of the system. In the case of a pure state $\mid \psi \rangle$, 
\begin{equation}
\label{eq:POVM}
P(m|\psi) = \mbox{Tr}\left(\hat{E}_m \hat{\rho} \right) = \langle \psi \mid \hat{E}_m \mid \psi \rangle.
\end{equation}
Thus, the statistical operator describes the causality relation between the input state $\mid \psi \rangle$ and a qualitative outcome $m$ observed as an effect of the system on the technical devices used in the measurement. Importantly, the notion of a physical property does not enter into this description at all, despite the fact that the measurement operator is a summary of dynamics that are described in terms of quantitative physical properties of the system. To recover the original connection between physical properties and measurement, it is necessary to introduce the notion of quantitative physical properties in the form of self-adjoint operators $\hat{A}$. In most introductions to quantum mechanics, these operators are used to establish a somewhat unclear analogy with classical physics by suggesting that they could somehow replace the classical notion of numbers. A more specific explanation of the role of operators would be that they describe a combination of quantity and quality by mathematically combining eigenstates $\mid a \rangle$ associated with precise measurement results with eigenvalues that represent the quantitative outcomes $A_a$ observed in such measurements. The self-adjoint operator itself can then be expressed by its spectral decomposition as
\begin{equation}
\hat{A} = \sum_a A_a \mid a \rangle \langle a \mid.
\end{equation}
Quite possibly, some confusion is created by the fact that many textbook introductions seem to make the implicit assumptions that energy is so fundamental to physics that neither the Hamilton operator nor any of the operators that appear in its quantitative definitions need any further explanation. In the light of the recent controversies on quantum measurement, it may be good to ask whether this implicit assumption is really justified. In general, the operator $\hat{A}$ may appear as part of a Hamiltonian, or it may be a generator of the system dynamics in its own right. However, it is not clear whether we can observe this role of the operator experimentally, and it is not clear whether this role is similar to the role of a corresponding classical quantity in the dynamics of a system \cite{Hof14,Hof16}. It may therefore be helpful to focus first on the physical meaning of the operator in the context of a quantum measurement with a well-defined observable outcome. 

As mentioned above, the formulation of the operator itself combines a potential measurement outcome $\mid a \rangle$ with a value $A_a$ that describes the magnitude of an effect caused by the physical property $\hat{A}$. However, the value $A_a$ does not enter into the theoretical description of such an ideal measurement of $\hat{A}$. It is sufficient to identify the individual outcomes $a$, which are obtained with a probability given by  
\begin{equation}
P(a|\psi) = \langle \psi \mid \hat{E}_a \mid \psi \rangle = |\langle a | \psi \rangle|^2.
\end{equation}
In this measurement, the quantity $\hat{A}$ is determined by assigning a measurement value of $A_a$ to each of the outcomes $a$. The only physical motivation for the specific assignment of values $a$ is obtained from the interpretation of the dynamics that relate the system property $\hat{A}$ to the measurement outcomes $a$, and this interpretation is in turn based on the analysis of the interaction between the system and the meter in the Hilbert space formalism. 

In a more realistic description of measurements, it is rarely possible to identify each outcome with a precise value of $\hat{A}$. If the outcome $m$ is partially decided by the quantity $\hat{A}$, we can still associate an estimate $\tilde{A}_m$ of the quantity $\hat{A}$ with each outcome. Since this is a quantitative estimate, its precision cannot be given in qualitative terms. Specifically, it makes no sense to ask for a probability that the estimate is correct. Instead, the magnitude of the error must be quantified in terms of the difference between the estimate $\tilde{A}_m$ and the target observable $\hat{A}$. In the operator formalism, this quantitative definition of an error is expressed by
\begin{equation}
\label{eq:error}
\hat{\epsilon}_m(A) = \tilde{A}_m - \hat{A}. 
\end{equation}
Since this error is an operator, its evaluation requires an application of the Hilbert space algebra. In cases where the measurement operator $\hat{E}_m$ does not commute with the operator $\hat{\epsilon}_m(A)$ describing its error, this evaluation contains non-classical correlations between the measurement outcome $m$ and the observable $\hat{A}$ \cite{Iin16}. The quantitative relation between the outcome $m$ and the physical property $\hat{A}$ is therefore a non-trivial scientific problem that cannot be reduced to the classical notions of information or statistics. In the following, we will address this problem by analyzing  the Hilbert space expression of quantitative errors given by Eq.(\ref{eq:error}) with respect to the quantitative relation they establish between the general measurement outcomes $m$ and the precise measurement outcomes $a$ associated with the eigenvalues $A_a$ of the physical property described by the operator $\hat{A}$.

\section{Error statistics in Hilbert space}
\label{sec:jsw}

Although there are many ways of evaluating statistical errors, a widely used measure is the variance, which is defined as the average square of the error values. In the context of quantum measurements, this error measure was first applied to a general Hilbert space description of measurement interactions by Ozawa \cite{Oza03}, who found that the operator ordering obtained from the detailed description of the measurement dynamics results in an error of
\begin{equation}
\label{eq:oz}
\varepsilon^2(A) = \sum_m \langle \psi \mid (\tilde{A}_m - \hat{A}) \hat{E}_m(\tilde{A}_m-\hat{A}) \mid \psi \rangle.
\end{equation}
This formulation of a measurement error has recently attracted a lot of attention since it provides a description of errors that seems to defy the uncertainty principle \cite{Has12,Roz12,Bae13,Wes13,Bra13,Rin14}. However, it is important to recognize that this result relies on the inclusion of information from the initial state $\mid \psi \rangle$, which is a separate condition that is not part of the theoretical description of the measurement process given by $\hat{E}_m$ \cite{Dre14,Hof03}. To understand the physics described by this quantification of the error, it is therefore important to consider how quantum mechanics expresses the relation between the physical property $\hat{A}$ and the observable causality that connects the initial conditions $\psi$ with the final outcomes $m$. 

What makes it difficult to understand the physics described by the mathematical form of Eq.(\ref{eq:oz}) is that it relates a qualitative outcome to an operator, thereby creating an artificial asymmetry in the description of the physical process. To restore the symmetry in terms of measurement outcomes, we can make use of the fact that the operator $\hat{A}$ is associated with a set of projective measurements $\hat{E}_a=\mid a \rangle \langle a \mid$ for which the eigenvalues $A_a$ represent an error free set of measurement outcomes. The error $\varepsilon^2(A)$ therefore evaluates a quantitative difference between the assignment of $\tilde{A}_m$ to $m$ and the assignment of $A_a$ to $a$. Importantly, Hilbert space describes a well defined correlation between these two assignments, even though the formalism also indicates that it is impossible to combine the outcomes $m$ and $a$ into a joint measurement corresponding to the logical combination of ``$m$ AND $a$''. Quantum theory thus provides a unique definition of non-classical correlations between the two quantitative evaluations, even though the outcomes $m$ and $a$ never share a joint reality. 

Mathematically, it is possible to trace back the non-classical correlations to joint statistical weights, since the contribution from the hypothetical outcome $a$ is exclusively associated with the value $A_a$, in the same way that the outcome $m$ is exclusively associated with its estimate $\tilde{A}_m$. Even though there exists no joint measurement of $a$ and $m$, it is therefore possible to define a joint statistical weight $P(a,m|\psi)$ for any combination of $a$ and $m$ by evaluating how the changes in the error $\varepsilon^2(A)$ caused by variations of $A_a$ and $\tilde{A}_m$ are correlated. Specifically, the joint statistical weight of $a$ and $m$ is obtained by
\begin{equation}
\label{eq:di}
P(a,m|\psi) = - \frac{1}{2} \frac{\partial^2}{\partial A_a \partial \tilde{A}_m} \varepsilon^2(A)
\end{equation}
We can thus derive a uniquely defined joint statistical weight of the measurement outcomes $a$ and $m$ from the error measure $\varepsilon^2(A)$ defined by the operators $\hat{A}$ and $\hat{E}_m$. Importantly, this joint statistical weight is different from a joint probability because it does not require a joint reality of $a$ and $m$. Instead, it is based entirely on the quantitative relation between the estimates of $\hat{A}$ obtained under the two different circumstances described by the two separate sets of outcomes $\{a\}$ and $\{m\}$. Since this quantitative relation is described by the Hilbert space formalism, the joint statistical weight can also be described in terms of states and operators. Using the Hilbert space expression of the measurement errors in Eq.(\ref{eq:oz}), the joint statistical weight describes a relation between the measurement operators $\hat{E}_m$ and $\mid a \rangle\langle a \mid$ in the state $\mid \psi \rangle$,
\begin{equation}
\label{eq:jsw}
P(a,m|\psi) = \mbox{Re}\left(\langle \psi \mid \hat{E}_m \mid a \rangle \langle a \mid \psi \rangle \right).
\end{equation}
Thus the quantitative evaluation of errors corresponds to joint statistical weights for measurement outcomes of measurements that cannot be performed jointly and therefore have no joint reality. This observation is particularly significant, because the mathematical form in Eq.(\ref{eq:jsw}) results in the assignment of non-positive joint statistical weights in several important cases that we will discuss in more detail in the following.

At this point, it may be worth noting that the relation between non-positive joint statistical weights and the measurement uncertainties introduced by Ozawa has been discussed previously in the context of quasi-probability representations of quantum statistics. Specifically, the joint statistical weights in Eq.(\ref{eq:jsw}) are the real parts of the Dirac distribution of the state $\mid \psi \rangle$, a quasi-probability representation that is closely associated with the algebra of weak values \cite{Hal04,Lun10,Dir45,Joh07,Lun12,Hof12}. It has also been shown that the measurement error can be observed experimentally by performing a weak measurement of $\hat{E}_a=\mid a \rangle\langle a \mid$ on the input state \cite{Lun10,Roz12}, and the present discussion was originally motivated by an attempt to understand the physics of these experimental results better \cite{Iin16}. Here, we would like to address the problem that such measurement results cannot be interpreted as a relative frequency of joint outcomes, since the individual weak measurements represent only a statistical preference for the outcome $a$ and not an exclusive selection of $a$ \cite{Dre14}. It is therefore necessary to consider the role of the outcomes $a$ in the experiments more carefully. As discussed above, the joint statistical weights actually describe quantitative relations between non-commuting physical properties in the absence of any joint reality. Based on a more thorough analysis of the mathematical relations, we have shown that the joint statistical weight is defined by the way that the quantum formalism attaches physical quantities to the outcomes of individual measurements. This means that the notion of a joint statistical weight does not require any joint realities of $a$ and $m$, which explains why its values can be negative. The present approach thus allows us to identify the relations between physical properties that are responsible for the appearance of negative joint statistical weights, which can help us understand the paradoxical nature of non-classical correlations \cite{Res04,Jor06,Tol07,Lun09,Yok09,Gog11,Suz12,Den14,Hof15}. 

Before we proceed to analyze the quantitative relations that cause the appearance of negative joint statistical weights, it may be useful to reflect a bit on the more familiar aspects of joint statistical weights and the reason why they are often confused with conventional joint probabilities, despite the fact that they can take negative values. In fact, we should not forget that the concept of ``uncertainty'' did not originate from quantum mechanics, but from practical considerations about the role of quantities in physics. The notion of a perfectly precise measurement only makes sense in the context of discrete values, which only emerge as a result of quantization. In classical physics, all physical properties are necessarily given by continuous and real valued quantities, so perfect precision would require the measurement of a never ending series of digits. Classical physics is therefore based on the idea that physical quantities can be used to describe physical situations even when the precision with which we can know these quantities is limited. This means that the physical properties only appear in the measurement outcomes $m$ because we can identify a quantitastive causality relation that connects the physical property $\hat{A}$ of an object with the outcome $m$, even though this relation cannot be described by a simple identity between the outcome $m$ and a hypothetical precise outcome $a$. In classical physics, this problem can be solved by the joint statistical weights defined in Eq.(\ref{eq:di}), where the joint statistical weight corresponds to a hypothetical joint probability of the outcome $m$ and the correct result $a$. However, quantum theory suggests that the assumption of such a joint probability is is a misinterpretation of the physics, even in the case where the joint statistical weights are positive. 

A close analogy to the classical statistical argument is obtained when the measurement operator $\hat{E}_m$ commutes with $\hat{A}$, so that the eigenstates $\mid a \rangle$ are also eigenstates of $\hat{E}_m$. In this case, it is tempting to argue that $\hat{E}_m$ describes a conventional conditional probability $P(m|a)$, given by the eigenvalue relation 
\begin{equation}
\hat{E}_m \mid a \rangle = P(m|a) \mid a \rangle.
\end{equation} 
As a result, the joint statistical weight of Eq.(\ref{eq:jsw}) can be written as
\begin{equation}
\label{eq:sequence}
P(a,m|\psi) = P(m|a)P(a|\psi),
\end{equation} 
which corresponds to the probability of measuring first $a$ and then $m$ in a sequence of two measurements. However, we should remember that this formula is used to relate $m$ to $a$ in the absence of a measurement of $a$. The actual role of the joint statistical weight $P(a,m|\psi)$ is to provide a universally valid description of the causality relation between $m$ and $a$ which identifies the influence of the physical quantity $\hat{A}$ on the outcome $m$. Thus, we should be careful to note that the hypothesis of a joint reality $(a,m)$ is not even needed when the observables commute, so that the positivity of the joint statistical weight $P(a,m|\psi)$ should not be misunderstood as an argument in favor of realist models.  

Interestingly, there exists one case where it is possible to identify a joint reality of $a$ and $m$, and this case does not even require commutativity of $\hat{E}_m$ and $\hat{A}$. If the initial state is an eigenstate of $\hat{A}$, the causality relation between $a$ and $m$ is given by the experimentally observed conditional probability 
\begin{equation}
P(m|a) = \langle a \mid \hat{E}_m \mid a \rangle. 
\end{equation} 
This expression shows that the outcome $m$ is sensitive to $a$, and therefore to the quantity $\hat{A}$ associated with $a$. However, it is not possible to apply $P(m|a)$ to any other input state $\mid \psi \rangle$, since Eq.(\ref{eq:sequence}) only applies when $\hat{E}_m$ and $\hat{A}$ commute. In the case of non-commuting operators, the consistency of quantum theory requires that the only statistical representation of the causality by which $\hat{A}$ causes the outcome $m$ is given by Eq.(\ref{eq:jsw}), where the unavoidable negative joint statistical weights represent the quantum modifications of causality by non-classical correlations between the physical properties. 

The conclusion from the above discussion is that it is insufficient to associate physical quantities $\hat{A}$ exclusively with the outcomes of precise measurements of the eigenvalue outcomes $a$. The role of physical properties $\hat{A}$ is not limited to the individual measurement outcomes $a$, since non-commuting physical properties are related to each other quantitatively by the operator algebra of Hilbert space. In the following, we will use the definition of quantitative errors in Eq.(\ref{eq:oz}) to identify precise quantitative relations between non-commuting physical properties.

\section{Error free measurement}
\label{sec:errorfree} 

Eq.(\ref{eq:oz}) defines the quantitative error of the estimate $\tilde{A}_m$ for a measurement outcome of $m$ for any set of measurement operators $\{ \hat{E}_m \}$. If this error is zero, we can conclude that the quantity $\hat{A}$ is precisely determined by the combination of initial conditions $\mid \psi \rangle$ and final outcomes $\hat{E}_m$. Error free measurements thus provide us with fully deterministic relations between the physical properties represented by these conditions, resulting in a quantitative description of measurement causality that explains the physics without any statistical concepts. 

To identify the conditions under which a measurement has zero error, we make use of the fact that Eq.(\ref{eq:oz}) describes the error as a sum of positive contributions from each measurement outcome $m$. Because of this, the total error can only be zero if the contribution for every $m$ is itself zero. We therefore find that the necessary and sufficient condition for an error of $\varepsilon^2(A)=0$ is that every estimate $\tilde{A}_m$ must satisfy the relation
\begin{equation}
\langle \psi \mid (\tilde{A}_m - \hat{A})\hat{E}_m(\tilde{A}_m - \hat{A}) \mid \psi \rangle= 0.
\end{equation}
Since each operator $\hat{E}_m$ can be replaced by its spectral decomposition, a value of zero is only obtained when $\hat{E}_m$ can be expressed by a single pure state projector, or when $\hat{E}_m$ is an arbitrary combination of pure state projectors that achieve an error of zero for the same value of $\tilde{A}_m$. We can therefore conclude that error free measurements can always be represented by measurement operators of the form $\hat{E}_m=\lambda \mid m \rangle \langle m \mid$. Using these measurement operators, the condition for error free measurements can be simplified to 
\begin{equation}
\label{eq:wv}
\tilde{A}_m = \frac{\langle m \mid \hat{A} \mid \psi \rangle}{\langle m \mid \psi \rangle}.
\end{equation}
Consistent with previous observations \cite{Hos10,Hof11a}, the necessary and sufficient condition for an error free measurement according to Eq.(\ref{eq:oz}) is that the estimates $\tilde{A}_m$ are all equal to an expression of the Hilbert space algebra that is usually identified as the weak value of $\hat{A}$ for the state $\mid \psi \rangle$ post-selected on $\mid m \rangle$. Importantly, this result is independent of the actual performance of weak measurements and demonstrates that weak values are part of the standard operator formalism. In an error free measurement, the physical meaning of $\tilde{A}_m$ is that of a precise quantitative evaluation of $\hat{A}$ associated with every measurement outcome $m$. Note that no post-selection is considered. Instead, the spectrum of values of $\tilde{A}_m$ assigned to the different outcomes $m$ replaces the eigenvalue spectrum of $\hat{A}$ as an equally valid representation of the statistics of $\hat{A}$ in the state $\mid \psi \rangle$.

In principle, Eq.(\ref{eq:wv}) indicates that any projective measurement $\{\mid m \rangle\}$ can have a measurement error of $\varepsilon^2(A)=0$ if the corresponding set of weak values of $m$ are chosen for the estimates $\tilde{A}_m$. However, weak values are generally complex, and the imaginary part is usually obtained in a dynamical response of the system that is not directly connected to the quantity represented by $\hat{A}$ \cite{Hof11b,Dre12}. Since $\tilde{A}_m$ is an estimate of the quantity $\hat{A}$, it does not have an imaginary part and Eq.(\ref{eq:wv}) can only be satisfied if the weak value is real from the start. Effectively, the condition for an error free measurement is that $\hat{A}$, $\{\mid m \rangle \}$ and $\mid \psi \rangle$ must satisfy the relations
\begin{equation}
\label{eq:Acond}
\mbox{Im}\left(\frac{\langle m \mid \hat{A} \mid \psi \rangle}{\langle m \mid \psi \rangle}\right)=0
\end{equation}
for all outcomes $m$. It should be noted that this still leaves a wide variety of cases where neither $\mid \psi \rangle$ nor any of the $\mid m \rangle$ are eigenstates of $\hat{A}$. In particular, Eq.(\ref{eq:Acond}) will be satisfied whenever both $\mid \psi \rangle$ and $\mid m \rangle$ can be expressed by superpositions of eigenstates $\mid a \rangle$ with real number coefficients. It is thus possible to find examples of error free measurements by constructing a basis $\{\mid m \rangle\}$ from real valued superpositions of $\{\mid a \rangle\}$ and choosing a non-orthogonal real valued superposition as input state. 

The possibility of error free measurements $\hat{E}_m$ that do not commute with the target observable $\hat{A}$ indicates that there is a precise quantitative relation between $m$, $\hat{A}$, and $\psi$ which is expressed by the error free measurement result $\tilde{A}_m$. In the Hilbert space formalism, this quantitative relation can be described as a relation between three operators, where one operator is associated with the measurement outcomes $m$ and another operator is associated with the initial state $\mid \psi \rangle$. Specifically, the physical quantity $\hat{A}$ can be expressed as a sum of a physical quantity $\hat{B}$ that has a value of $B_\psi$ for the initial state $\mid \psi \rangle$ and a physical quantity $\hat{M}$ that has a value of $M_m$ for the measurement outcome $m$. The quantitative relation between $\hat{A}$ and the measurement outcomes is then given by the operator sum
\begin{equation}
\label{eq:add}
\hat{A} = \hat{B} + \hat{M}.
\end{equation}
This equation represents the additivity of physical properties in the Hilbert space formalism. In the present context, it is important that each of the three operators has a completely different set of eigenstates. The initial state $\mid \psi \rangle$ is an eigenstate of the operator $\hat{B}$, which means that the physical property $\hat{B}$ of the system was determined as a result of the initial quantum state preparation. The operator $\hat{M}$ represents a quantity determined exclusively by the measurement outcomes, so its eigenstates are given by the measurement basis $\{\mid m \rangle\}$, which are different from the eigenstates $\{\mid a \rangle\}$ of $\hat{A}$. Since $\hat{M}$ describes the dependence of $\tilde{A}_m$ on $m$, it is given by
\begin{equation}
\label{eq:Mform}
\hat{M} = \sum_m \;\; \left(\tilde{A}_m - B_\psi \right) \;\;\mid m \rangle \langle m \mid.
\end{equation}
Importantly, the measurement outcome $m$ is completely explained by the physical property $\hat{M}$ of the system. The fact that $\{\mid m \rangle\}$ is a complete orthogonal basis of the Hilbert space of the system means that the properties of the meter system have no effect on the measurement outcome. The measurement outcome is an objective consequence of the value $M_m$ of the physical property $\hat{M}$. 

The relation between the two non-commuting physical quantities $\hat{M}$ and $\hat{A}$ can only be established through the third physical quantity $\hat{B}$ associated with the initial state $\mid \psi \rangle$. This physical property has a fixed value of $B_\psi$ for its eigenstate $\mid \psi \rangle$, and this value contributes equally to each estimate $\tilde{A}_m$. It is therefore not possible to identify a unique value of $B_\psi$. The reason for this ambiguity is that $\hat{M}$ and $\hat{B}$ are only distinguished by their different eigenstates, not by their absolute values. Adding or subtracting a constant value does not change the eigenstates of the operators, so it is not possible to decide from mathematics alone how the average contribution to $\hat{A}$ should be split between $\hat{M}$ and $\hat{B}$. However, it is interesting to note that Eq.(\ref{eq:add}) automatically determines the complete set of eigenstates of $\hat{B}$, since the operator $\hat{B}$ is determined by the difference of $\hat{A}$ and $\hat{M}$. By using this definition of the operator $\hat{B}$, we can confirm that $B_\psi$ is the eigenvalue of $\hat{B}$ for the initial state $\mid \psi \rangle$,
\begin{eqnarray}
\hat{B} \mid \psi \rangle &=& (\hat{A} - \hat{M}) \mid \psi \rangle
\nonumber \\ &=& \sum_m \mid m \rangle \langle m \mid (\hat{A}-\tilde{A}_m + B_\psi) \mid \psi \rangle
\nonumber \\ &=& B_\psi \mid \psi \rangle.
\end{eqnarray}
Note that operator expression $\hat{A}-\tilde{A}_m$ in the second line can be identified with the error operator $\hat{\epsilon}_m(A)$ in Eq.(\ref{eq:error}). Thus, the disappearance of the error ensures that $\mid \psi \rangle$ is an eigenstate of the difference between $\hat{A}$ and $\hat{M}$. 

We can now explain why a quantitative measurement error of zero does not require a joint reality of eigenstates, and why it is usually obtained for measured values $\tilde{A}_m$ that are different from the eigenvalues $A_a$ of the target observable $\hat{A}$. The quantitative relation that makes this error free estimate possible is the additivity of physical properties, as expressed by the operator sum in Eq.(\ref{eq:add}). The quantitative relation between physical properties expressed by such operator sums cannot be explained in terms of the eigenvalues of the operators, because the sum of the eigenvalues is different from the eigenvalues of the sum. It is this non-classical quantitative relation between physical properties that is expressed by the error free weak value estimates of Eq.(\ref{eq:wv}). In particular, the initial state $\mid \psi \rangle$ always corresponds to a known physical property $\hat{B}$ that evaluates the quantitative difference $(\hat{A}-\hat{M})$ between the target observable $\hat{A}$ and the measured quantity $\hat{M}$.

The explanation of error free measurements by precise quantitative relations between physical properties shows that it is not sufficient to identify physical properties with the particular set of idealized measurement outcomes represented by their eigenstates. In many circumstances, a more meaningful definition of a physical property will involve a combination of initial and final conditions, and in these cases, the eigenstate decomposition of the operator will have no physical meaning. A particularly striking example of this problem is given by the description of time evolution in quantum mechanics, where it is reasonable to be interested in the rate of change for a specific observable. Such a rate of change is usually observed by a sequence of measurements, where the first measurement prepares the initial value and the final measurement determines the final value. This corresponds exactly to the situation discussed above, where $\hat{M}$ is simply $\hat{B}(t)$ and the rate of change is given by $\hat{A}/t = (\hat{B}(t) - \hat{B}(0))/t$. Importantly, the time derivative of an operator does not commute with that operator. This means that the classical concept of trajectories must be replaced with quantitative operator relations that describe the observable effects of the time evolution. For the motion of a particle, it is only possible to assign an error free position between initial preparation and final measurement if the weak value of position associated with the initial and final conditions is a real value, and even then it is not correct to identify the value of the position operator with the detection of a particle at that position, since the quantitative estimate of position is different from an estimate of the probability of finding a particle at that position. However, it should be kept in mind that positions are usually observed in the quantitative dependence of interaction strength on distance, so the quantitative evaluation of position is closer to the observable physics than the qualitative assignment of position associated with particle detections. 

Quantitative relations are at the heart of the definitions of physical concepts. It is therefore not possible to reduce the physics of a system to the measurement outcomes observed in limited sets of specific measurements. Instead, it is necessary to explain how physical quantities appear in different contexts, each of which is represented by a specific combination of initial and final conditions. Problems arise because the quantities observed in each context are different from each other. For this reason, it is extremely important to examine the relation between different measurement outcomes described by the joint statistical weights given by Eq. (\ref{eq:jsw}) in more detail.

\section{Relations between physical properties}
\label{sec:context}

As shown in sec. \ref{sec:jsw}, it is possible to express the error statistics of the measurement $\hat{E}_m$ in terms of the joint statistical weights of $a$ and $m$ defined by Eq.(\ref{eq:di}). The measurement error $\varepsilon^2(A)$ itself is then given by
\begin{equation}
\label{eq:epstats}
\varepsilon^2(A) = \sum_{m,a}(\tilde{A}_m-A_a)^2 \; P(a,m|\psi),
\end{equation}
an expression that corresponds to classical error statistics, where $P(a,m|\psi)$ would be the probability of a joint outcome of $a$ and $m$. However, quantum mechanics does not permit an interpretation of Eq.(\ref{eq:epstats}) in terms of joint realities of $a$ and $m$, as indicated by the possibility of negative joint statistical weights. 

As shown in the previous section, the error $\varepsilon^2(A)$ can be zero in situations where the measured values $\tilde{A}_m$ are generally different from the eigenvalues $A_a$. Since this means that the squared differences between the two values are all positive and non-zero, the sum in Eq.(\ref{eq:epstats}) can only be zero if some of the joint statistical weights $P(a,m|\psi)$ are negative. We can therefore conclude from the discussions above that negative joint statistical weights are a necessary characteristic of error free measurements. Since the error free measurement can also be explained in terms of the quantitative relations between the physical quantities represented by Hilbert space operators, these non-positive statistical weights are simply a more specific formulation of the physics represented by the established formalism. In particular, non-positive joint statistical weights describe the necessary relations between the measurement outcomes $m$ and the measurement outcomes $a$ that are required in order to reconcile the quantitative relations between the operators $\hat{A}$, $\hat{M}$ and $\hat{B}$ with the eigenvalues of the respective operators. We can therefore combine the results for the joint statistical weights with the results of the previous section to arrive at a better understanding of how joint statistical weights describe the relations between quantities and measurement outcomes in different measurement contexts.

To make the connection between error free measurements and joint statistical weights, we first derive the optimal estimate $\tilde{A}_m$ for a given joint statistical weight $P(a,m|\psi)$ by minimizing the error as given by Eq.(\ref{eq:epstats}). As expected from the formal analogy with classical error statistics, the minimal error is obtained when the estimates $\tilde{A}_m$ correspond to conditional averages of the joint statistical weights,
\begin{equation}
\label{eq:optest}
\tilde{A}_m =  \sum_{a} A_a \frac{P(a,m|\psi)}{P(m|\psi)}.
\end{equation}
Now, we need to identify the condition under which this optimized estimate is error free. For this purpose, we can reformulate (\ref{eq:wv}) by using the spectral decomposition of the operator $\hat{A}$ to arrive at a condition for error free measurements that relates two sums over $a$,
\begin{equation}
\label{eq:condition}
\sum_{a} A_a \frac{P(a,m|\psi)}{P(m|\psi)} = \sum_{a} A_a \frac{\langle \psi \mid m \rangle \langle m \mid a \rangle \langle a \mid \psi \rangle}{\langle \psi \mid m \rangle \langle m \mid \psi \rangle}.
\end{equation}
This condition is both necessary and sufficient for error free measurements. However, it may be worth noting that a sufficient condition can be obtained by demanding equality for every element of the sum. In this case, the conditions for error free measurements do not depend on the eigenvalues $A_a$, so it applies equally to measurements of any non-linear function of $\hat{A}$. This condition is given by 
\begin{equation}
\label{eq:shortcond}
P(a,m|\psi) =  \langle \psi \mid m \rangle \langle m \mid a \rangle \langle a \mid \psi \rangle.
\end{equation}
Comparison with Eq.(\ref{eq:jsw}) shows that this equality is satisfied whenever the right hand side is a real number,
\begin{equation}
\label{eq:Imcond}
\mbox{Im}\left(  \langle \psi \mid m \rangle \langle m \mid a \rangle \langle a \mid \psi \rangle \right) = 0.
\end{equation}
Whenever Eq.(\ref{eq:Imcond}) is satisfied for all combinations of $a$ and $m$, the measurement is error free and Eq.(\ref{eq:optest}) expresses the relation between the values of $\hat{A}$ obtained in the measurement $m$ and the values of $\hat{A}$ obtained in a measurement of $a$. In general, this condition restricts the possible choices of $\mid \psi \rangle$ for which joint error free measurements of $\hat{A}$ and $\hat{M}$ are possible. However, a large number of non-trivial solutions can be obtained by finding pairs of basis systems where all inner products $\langle m \mid a \rangle$ are real numbers. In that case, $\mid \psi \rangle$ can be any superposition of $\mid a \rangle$ or $\mid m \rangle$ with real valued coefficients, as suggested below Eq.(\ref{eq:Acond}). It may also be interesting to note that the right hand side of Eq.(\ref{eq:shortcond}) is the Dirac distribution of the state $\mid \psi \rangle$. We thus find that a measurement is error free whenever the Dirac distribution of the measurement outcomes $m$ and the outcomes of a projective measurement $a$ takes only real values.

We have now confirmed that Eq.(\ref{eq:optest}) describes the relation between an error free estimate $\tilde{A}_m$ and the eigenvalues $A_a$ in terms of the corresponding joint statistical weights $P(a,m|\psi)$. We also know from the discussion in Sec. \ref{sec:errorfree} that $\tilde{A}_m$ is error free because it can be obtained from the operator relation $\hat{A}=\hat{M}+\hat{B}$ as a sum of eigenvalues of $\hat{M}$ and $\hat{B}$,
\begin{equation}
\label{eq:eigenest}
\tilde{A}_m = M_m + B_\psi.
\end{equation}
We can now address the problem that the quantitative relation between the operators $\hat{A}$ and $\hat{M}$, $\hat{B}$ does not correspond to any quantitative relation between the eigenvalues $A_a$ and $M_m$, $B_\psi$. According to Eq.(\ref{eq:optest}) and Eq.(\ref{eq:eigenest}), the relation between the eigenvalues of the operators is given by 
\begin{equation}
\label{eq:Mm}
M_m =  \sum_{a} (A_a - B_\psi) \frac{P(a,m|\psi)}{P(m|\psi)}.
\end{equation}
Here, the joint statistical weights $P(a,m|\psi)$ describe a transformation procedure by which the differences of the eigenvalues of $\hat{A}$ and $\hat{B}$ are converted into an eigenvalue of $\hat{M}$. This transformation highlights the fact that the measurement of $m$ and the measurement of $a$ are incompatible contexts, whereas the operator relations are context independent descriptions of the physical object. Therefore the quantitative relation between eigenvalues observed in the different contexts need to be described by transormations between these different measurement contexts. 

It is important to realize that we can only identify a physical object if its properties are independent of the context - otherwise, we would not be able to tell whether we are really observing the same system in different experimental situations. It seems that most arguments about incompatibilities of measurement contexts in quantum mechanics overlook this important requirement. Contextuality is only acceptable if we know precisely how the relation between the contexts can be explained in terms of objective physical properties. The discussion above shows that the Hilbert space formalism offers a solution to this problem once we realize that the operators provide a context independent definition of quantities that includes a precise prescription for the transformation between different contexts. 

It should also be noted that the introduction of $\hat{M}$ establishes a completely symmetric relation between a measurement of $m$ and a measurement of $a$. This means that a measurement of $a$ is not only error free in $\hat{A}$, but also error free in $\hat{M}$, with the error free estimate of $\hat{M}$ given by 
\begin{equation}
\label{eq:Moptest}
\tilde{M}_a =  \sum_{m} M_m \frac{P(a,m|\psi)}{P(a|\psi)}.
\end{equation}
From this relation, we can derive the reverse transformation that converts the eigenvalues $M_m$ of $\hat{M}$ back into eigenvalues $A_a$ of $\hat{A}$,
\begin{equation}
\label{eq:Aa}
A_a = \sum_{m} (M_m + B_\psi) \frac{P(a,m|\psi)}{P(a|\psi)}.
\end{equation}
Here, the eigenvalues represent the quantitative outcomes obtained in the specific context of a fully resolved measurement of the corresponding quantity. To explain the quantitative relation between the values $A_a$ obtained in the context of outcomes $a$ and the values $M_m$ obtained in the context of outcomes $m$, non-positive joint statistical weights can be used to convert the two sets of values into each other. 

Although there is a formal analogy between this transformation and the calculation of conditional averages, it is important to remember that $\psi$ represents the physical property $\hat{B}$ in the operator relation given by Eq.(\ref{eq:add}). Therefore the joint statistical weight is a relation between physical properties that does not involve any specific physical situation. Likewise, the quantities related to each other by the joint statistical weights are eigenvalues and not conditional averages. It needs to be emphasized that the non-positive statistical weights are an expression of deterministic relations between the physical properties, equivalent in meaning to the original operator relation in Eq.(\ref{eq:add}), but more specific in their identification of the measurement contexts associated with the respective eigenvalues of the operators. 

The transformations discussed in this section show that the deterministic relations between the physical quantities $\hat{A}$, $\hat{M}$ and $\hat{B}$ can only be expressed in terms of their eigenvalues if the faulty notion of a joint reality is replaced by the more appropriate transformation between the measurement contexts described by the joint statistical weights derived in sec. \ref{sec:jsw}. Our analysis therefore clarifies the role of the measurement context in quantum measurements. Importantly, the measurement context can be defined by objective physical properties without any direct reference to the state of the environment or the measurement setup. The absence of a joint reality for the different contexts can then be explained completely in terms of the non-classical relations between the physical quantities of quantum systems.

\section{The origin of non-classical correlations}
\label{sec:ncc}

Perhaps the most confusing aspect of quantum mechanics is the rather peculiar combination of statistical concepts and deterministic quantitative relation in the mathematical formalism. As we have shown in the previous section, this results in the definition of joint statistical weights that appear to be similar to probability distributions but describe reversible transformations between two different measurement contexts. A direct statistical interpretation is impossible, since some of the joint statistical weights are necessarily negative. However, the error free assignment of measurement values $(\tilde{A}_m, M_m)$ to an outcome $m$, and of $(A_a, \tilde{M}_a)$ to an outcome $a$ indicates that statistical correlation between the quantities $\hat{A}$ and $\hat{M}$ in the initial state $\mid \psi \rangle$ can be determined from either one of the two measurements. Specifically, we can express the correlations by the statistical average over the product of the two error free values. The reason why this average product is independent of the measurement context is that the two averages are related by the transformation between contexts expressed by the joint statistical weight $P(a,m|\psi)$. It is therefore possible to express the average product using only the eigenvalues and the joint statistical weights. This context independent representation of non-classical correlations corresponds directly with the Hilbert space expression of the expectation value of the operator product,
\begin{eqnarray}
\label{eq:contextcorr}
C(A M|\psi) &=& \sum_m \tilde{A}_m M_m P(m|\psi)
\nonumber \\
&=& \sum_a A_a \tilde{M}_a P(a|\psi)
\nonumber \\
&=& \sum_{a,m} A_a M_m P(a,m|\psi) 
\nonumber \\
&=& \langle \psi \mid \hat{M}\hat{A} \mid \psi \rangle.
\end{eqnarray}
We thus find that the correlation defined by the operator product of $\hat{A}$ and $\hat{M}$ can be evaluated as product of the eigenvalues using the joint statistical weight, or as the statistical average of the product of error free estimates within any error free measurement context. In effect, the operator product of the two non-commuting operators $\hat{A}$ and $\hat{M}$ correctly describes the non-classical correlations observed in any measurement context that permits error free estimates of both $\hat{A}$ and $\hat{M}$. Here too, the joint statistical weight provides the correct description of the relation between the two contexts defined by the eigenstates. 

The probability distributions $P(m|\psi)$ and $P(a|\psi)$ describe the randomness of the initial condition $\psi$, which only determines the value of $\hat{B}$, but not the values of $\hat{A}$ or $\hat{M}$. The quantitative relation between the three properties means that the product average can be expressed in terms of the statistics of $\hat{A}$ or of $\hat{M}$ by using the fact that $\psi$ determines the value of $\hat{B}$,
\begin{eqnarray}
C(A M|\psi) &=& \langle \psi \mid \hat{A}^2 \mid \psi \rangle - B_\psi \langle \psi \mid \hat{A} \mid \psi \rangle
\nonumber \\
&=& \langle \psi \mid \hat{M}^2 \mid \psi \rangle + B_\psi \langle \psi \mid \hat{M} \mid \psi \rangle.
\end{eqnarray}
Thus, the non-classical correlations between $\hat{A}$ and $\hat{M}$ are fully determined by the fluctuations of measurement outcomes, whether the measurement is a measurement of $a$ or a measurement of $m$. 

For a specific measurement context, the correlations and the fluctuations are related by the appropriate quantitative relations between the error free estimate and the eigenvalues. For example, the statistics of correlations and fluctuations for the measurement outcomes $m$ is given by
\begin{eqnarray}
\label{eq:convert}
C(A M|\psi) &=& \sum_m \tilde{A}_m M_m P(m|\psi)
\nonumber \\
&=& \sum_m (M_m + B_\psi) M_m P(m|\psi)
\nonumber \\
&=&  \sum_m \tilde{A}_m (\tilde{A}_m-B_\psi) P(m|\psi).
\end{eqnarray}
Thus non-classical correlations can be traced back to the deterministic relations between the physical quantities $\hat{A}$ and $\hat{M}$ obtained for a fixed value of the quantity $\hat{B}$. The error free values $\tilde{A}_m$ indicate that the fluctuations in $\hat{A}$ and the fluctuations in $\hat{M}$ originate from the same source of randomness. The Hilbert space formalism describes this source of randomness in terms of the state vector $\mid \psi \rangle$, but the physical meaning is better explained by the transformation relations between the measurement contexts $a$ and $m$. 

In summary, the non-classical correlations expressed by products of non-commuting operators can be explained in terms of the transformations between measurement contexts given by the joint statistical weights. These correlations are directly observed in error free measurements, where the known relations between target observable and the measurement result provide a precise value of the target observable that is different from an eigenvalue. Nevertheless, the same non-classical correlations can be observed in all measurement contexts, demonstrating that it is the quantitative relations between physical properties and not the individual outcomes that are objectively real. 

\section{Conclusions}
\label{sec:concl}

Our quantitative analysis of measurement errors shows that the mathematical description of physical quantities by Hilbert space operators is justified by the causality relations between the physical properties that characterize the initial state and the physical properties associated with the outcomes of the final measurement. The additivity of physical quantities expressed by Eq.(\ref{eq:add}) has important consequences for the relations between different measurement contexts, since it provides a precise description of the transformations of physical quantities between the two contexts. Significantly, the relations between quantities and hence the relations between contexts are fully deterministic and represent the universal relations between physical properties that define a physical object. It is therefore possible to reconcile the different measurement concept with the identification of objective properties as the fundamental causes of measurement outcomes. 

The results presented here suggest that a purely information theoretic or statistical approach to quantum physics may be insufficient. The quantitative relations between physical properties cannot be described in terms of conventional probability distributions, because the relations between the values observed in different measurement contexts make it necessary to assign non-positive joint statistical weights. Oppositely, it is possible to derive a more detailed description of non-classical correlations from the relations between the physical quantities. Quantitative relations between physical properties are therefore more fundamental to quantum physics than the statistical properties associated with quantum states. 

Most importantly, this investigation into the relation between measurement outcomes and physical properties shows that all different measurement contexts relate to the same set of physical properties, so that the complete Hilbert space structure can be explained in terms of experimentally accessible physics. This kind of analysis should be applied to all quantum experiments, since it may well be the only way to establish a proper empirical foundation of quantum physics. At present, it is a weakness of quantum research that experimental results are only used to verify one specific aspect of the theory, which means that we deprive ourselves of more general insights into the actual physics. Here, we have shown that the quantitative relations between physical properties represent the actual physics of a quantum system, making it possible to relate different measurement contexts to each other. We believe that this result will be absolutely essential for future progress towards a more complete understanding of quantum physics.

\end{document}